\documentclass[pageno]{jpaper}

\usepackage[normalem]{ulem}

\newif\ifisdraft

\usepackage{times}
\usepackage[T1]{fontenc}
\usepackage[utf8]{inputenc}

\usepackage{amsthm}
\usepackage{scalerel, amssymb}
\usepackage{amsmath}

\usepackage{listings}

\usepackage{hyperref}
\usepackage{pdfpages}
\usepackage{titlesec}
\usepackage{xspace}

\usepackage[framemethod=default]{mdframed}

\mdfdefinestyle{inv}{%
	linecolor=black,linewidth=2pt,%
	topline=false,
	leftline=false,
	innertopmargin=1pt,
	skipbelow=\topskip, 
	skipabove=\baselineskip,
}

\mdfdefinestyle{defi}{%
	linecolor=black,linewidth=2pt,%
	frametitlerule=true,%
	frametitlebackgroundcolor=gray!20,
	innertopmargin=3pt,
	skipbelow=\topskip, 
	skipabove=\baselineskip,
}

\mdtheorem[style=defi]{defi}{Definition}
\mdtheorem[style=defi]{op}{Operation}
\mdtheorem[style=inv]{inv}{Invariant}

\usepackage{tikz}
\usetikzlibrary{calc}
\usetikzlibrary{automata,positioning}
\tikzset{every node/.append style={font=\small}}
\usepackage[all,cmtip]{xy}

\usepackage{color}
\definecolor{bluekeywords}{rgb}{0.13,0.13,1}
\definecolor{greencomments}{rgb}{0,0.5,0}
\definecolor{redstrings}{rgb}{0.9,0,0}

\lstdefinestyle{code}
{language=C,
	basicstyle=\tiny,
	showspaces=false,
	showtabs=false,
	breaklines=true,
	showstringspaces=false,
	breakatwhitespace=true,
	escapeinside={(*@}{@*)},
	commentstyle=\color{greencomments},
	keywordstyle=\color{bluekeywords}\bfseries,
	stringstyle=\color{redstrings},
	basicstyle=\ttfamily,
	morekeywords={uint8_t, uint16_t, uint32_t, uint64_t, size_t, err_t, regionid_t},}

\ifisdraft
\newcommand{\redcomment}[1]{\textcolor{red}{\textbf{#1}} }

\else
\newcommand{\redcomment}[1]{ }

\fi

\newcommand{\RH}[1]{\redcomment{RH: #1}}

\newcommand{\RA}[1]{\redcomment{RA: #1}}

\newcommand{\new}[1]{#1}

\newcommand{\ei}{{\it i)}\xspace}
\newcommand{\eii}{{\it ii)}\xspace}
\newcommand{\eiii}{{\it iii)}\xspace}

\newcommand{\isempty}[3]{%
    \if\relax\detokenize{#1}\relax
    #2%
    \else
    #3%
    \fi}

\begin{document}

\newcommand{\system}{{CleanQ}\xspace}
\newcommand{\System}{{CleanQ}\xspace}
\newcommand{\os}{\textit{AnonOS}\xspace}
\newcommand{\Os}{\textit{AnonOS}\xspace}

\title{\System: a lightweight, uniform, formally specified interface for 
intra-machine data transfer}

\author{Roni Haecki, Lukas Humbel, Reto Achermann, David Cock, Daniel Schwyn, Timothy Roscoe\\
	Systems Group, Department of Computer Science, ETH Zurich}

\date{}
\maketitle

\thispagestyle{empty}

 \begin{abstract}
We present \System, a high-performance operating-system interface for
descriptor-based data transfer with rigorous formal semantics, based on a
simple, formally-verified notion of \emph{ownership transfer}, with
a fast reference implementation.

\System aims to replace the current proliferation of similar, but
subtly diverse, and loosely specified, descriptor-based interfaces in
OS kernels and device drivers.   \System has strict semantics that not
only clarify both the implementation of the interface for different
hardware devices and software use-cases, but also enable composition of
modules as in more heavyweight frameworks like Unix streams. 

We motivate \System by showing that loose specifications derived from
implementation lead to security and correctness bugs in
production systems that a clean, formal, and easily-understandable
abstraction helps eliminate.  We further demonstrate by experiment
that there is negligible performance cost for a clean design: we show
overheads in the tens of cycles for operations, and comparable
end-to-end performance to the highly-tuned Virtio and DPDK implementations on
Linux.
\end{abstract}


\section{Introduction}
\label{sec:introduction}

\System is a uniform operating system interface for
transferring bulk data, which abstracts from and unifies a wide
variety of descriptor-based data transfer interfaces used by both
software and hardware in a modern OS.  

Queues based on descriptor rings are pervasive in OS code for moving
data between processes, hardware devices like network adaptors, kernel
and user space, etc.  Despite this there is a wide range of different
queue interfaces and implementations (even within a single OS).
So-called standard interfaces to queues, where they exist, are
typically specified informally using a reference implementation in C.

This leads to a serious problem, which we elaborate on in
Section~\ref{sec:background}.  Implementations cannot be reused (since
the semantics are subtly different), and thus implementation
\emph{bugs can recur} (and do) when a new queue is built.  The lack of
clear semantics mean that bugs also arise due to \emph{inappropriate
  use} of a given descriptor queue, by a client programmer who may not
understand its subtleties, and also make it
\emph{hard to compose} code modules which operate on queues of data,
as is possible with Unix Streams or other I/O frameworks.  Without a
formally sound description of a queue's behavior, it is
\emph{impossible to reason} about the correct behavior of the OS which
uses them.  Finally, a lack of uniformity is a lost opportunity
to \emph{build and apply standard tools} for debugging, validating,
profiling, and monitoring such queues at runtime, making OS
development more difficult and time-consuming.

\System addresses these problems not by simply proposing yet another
queue interface, but starting from a \emph{provably sound formal
specification} (presented in Section~\ref{sec:design}) of how a descriptor 
queue should behave. This gives
clear guarantees to clients of a queue (whether it be a device driver
or an inter-process communication system), and also states clear
obligations on the code that implements an end-point of the queue.

The specification is strict, and so the chance of subtle mismatches
between the expectations of client and implementation are eliminated.
It allows for full concurrency between actors, such as a driver process and a 
network card. It also subsumes the memory model in use: the client of a \System
queue does not need to be concerned about weak consistency or
non-coherent memory in order to write correct, portable code to use
it.  However, \System focuses purely on the data plane interface,
leaving flexibility to system designers in how such queues are
instantiated and provisioned.

From the specification, we then proceed to a C interface which
captures it.  This interface is highly general.  In
Section~\ref{sec:implementation} we describe it and demonstrate its
generality with a number of implementations we have built behind it,
for network cards, storage adaptors, and inter-process communication.
Moreover, the \System C interface composes: we also describe \System
modules that provide loopback, debugging and network stack functionality. 

Finally, we evaluate the overhead of using our \System implementation
modules on Linux and a microkernel-based research operating system in 
Section~\ref{sec:evaluation} to show both performance and portability.  
We show that, despite the strict semantics and highly specified, uniform interface,
\System is cheap: it is comparable to Virtio and DPDK in operation latency and
imposes less than 1\% overhead for set/get operations using Memcached
\cite{Fitzpatrick2004}.

\RH{I dont think this is necessary text (since we have a 11 page limit): 
	In the next section, we discuss the problematic nature of current
	descriptor-queue based interfaces in more detail.}



\section{Background and Motivation}
\label{sec:background}

\System is a formalization of descriptor rings. 
Rings of descriptors are a fairly pervasive technique for transferring
data between end-points (software processes or threads, GPUs, address
spaces, hardware I/O devices, virtual machines, etc.) in a modern OS.
Each descriptor refers to a region of memory (usually elsewhere) plus
some metadata, including which end of the communication ``owns'' the
data \new{(and metadata)}. The Linux kernel alone has at least 6 different descriptor
queue implementations, not including hardware-specific I/O queues for
network and storage devices. 

Descriptor rings work well because they highly decouple sender and
receiver: sending data between a user process and a high-performance
network adaptor using Intel's DPDK~\cite{DPDK}, for example, doesn't
require either side to touch payload data as part of the transfer, and
in the common case requires no synchronization between sender and
receiver.  In such drivers, interrupts and hardware register access
is only used for coarse-grained synchronization at low load levels;
for the most part each side of the communication can proceed in
parallel without explicit coordination.

However, while the \emph{technique} of descriptor rings is almost universal,
there is little consensus on what a given ring should look like.  A
great number of software interfaces, libraries and `standards' have
been proposed over time
(e.g.~\cite{Druschel:1993:FHC:168619.168634,Friedley2013,Russell2008,VMI,NetlinkMmap,Han:2012:MNP:2387880.2387894,Bjorling:2013:LBI})
all of which are variations or enhancements of the same basic theme.
Moreover, every new high-performance I/O device adopts a different
descriptor format for its I/O queues, including devices from the same
vendor
(e.g.~\cite{rtl8029_datasheet,intel82571_datasheet,intel82576_datasheet,intel82599_datasheet,intel631_manual,broadcom_netex2}).

\subsection{The Continuing Emergence of Descriptor Bugs}

This proliferation of implementations is often for good reasons: queue
implementations (whether communicating between processes or between
software and hardware devices) have different requirements in how they
are constructed and set up, metadata that may need to be passed with
each buffer, and additional, implementation-specific semantics
associated with enqueue and dequeue operations.  Examples are
Virtio~\cite{Russell2008} for buffer transfer between host and virtual 
machines, SKBufs\cite{skbufs} in the Linux network stack or mbufs in 
DPDK~\cite{DPDK}.
\RH{Reviewer suggest mentioning RDMA}
However, the proliferation of implementations combined with the
difficulty of getting memory semantics right leads to a steady stream
of serious bugs, ranging from performance problems to critical
security vulnerabilities.   

For example, the Virtio
framework used in the QEMU emulator and the KVM virtual
machine monitor has allowed a malicious guest to break security by
inserting more requests than the size of the
queue~\cite{CVE20165403}.   Changing the queue parameters in Virtio
has caused the hosting QEMU process to crash~\cite{CVE201717381}). 

Worse, these bugs have not only been appearing for a long time -- they
continue to appear~\cite{CVE20040555, CVE20081317, CVE20101187,
  CVE20159016, CVE20157613, CVE20151805, CVE20155336, CVE20165403,
  CVE201714916, CVE20177618, CVE201717381, CVE20179986} in Linux and
Android (and, we suspect, other system software).
\new{A new class of ``double-fetch'' bugs/vulnerabilities have
  recently appeared~\cite{Wang2017} whereby data is fetched twice but
changed by another party in between}. All these bugs
ultimately boil down to production code making incorrect assumptions
about when and how memory can be accessed safely by one side of a
descriptor queue-based channel: when it is safe to reuse a buffer,
when an endpoint can safely enqueue another buffer, etc. 

We argue the main reason this problem is not going away is the
prevalence of \emph{specification by implementation}: The
documentation--
where it exists at all--is written in English prose and not formally specified.
For example, consider the \texttt{virtqueue} mechanism in Virtio~\cite{Russell2008} for
transferring data between a device driver and a virtual device.
\begin{quote}
A \texttt{virtqueue} consists of (1) a \emph{descriptor table} specifying
which buffers a driver is using for its device, (2) an \emph{available
  ring} containing descriptors offered to the device, and (3) a
\emph{used ring} containing buffers processed by the device and
returned to the driver.
\end{quote}
While the specification makes it clear that
buffers available to the programmer are on one of the two rings, the 
corner-cases (such as adding a descriptor twice to a ring) are
undocumented.  Programmers are ultimately advised to read the code.  
To make things worse, Virtio has two different queue interfaces--one 
for the host and one for the guest--having slightly different semantics (e.g.
a \texttt{memcpy} on the host side).


\subsection{Lack of Portability and Reuse}

A further consequence of specification-via-implementation is
that correctness often depends on a particular memory model. 
The combination of program operations and fences/barriers required for
correct operation is not at all obvious from the documentation, and is
different for different processor architectures.   For example, when a
device driver enqueues a buffer, it first writes the buffer
contents, then the descriptor, and finally updates the head pointer of
the ring buffer to inform the device that there is a new
descriptor.  On a machine with a weak memory model, these
operations can be reordered and the device notified before the
descriptor is written to memory. 

Consider, for example, invoking a Virtio~\cite{Russell2008} or
fbuf~\cite{Druschel:1993:FHC:168619.168634} based queue between two cores: 
an Intel x86 machine implements Total Store
Ordering~\cite{Sewell2010}, leading to a relatively simple
implementation: if a thread performs two writes $w1$ before $w2$,
another thread that observes $w2$ will also have observed the effect of $w1$.  
Implementing buffer transfer for x86-TSO or similar models, it suffices
to ensure correct write ordering.

However, machines with weak memory consistency
like ARMv7, ARMv8, or IBM Power~\cite{Sewell2016} allow considerable
relaxation in the visibility order from a given core: any load, store,
\emph{or} atomic instruction can be extensively reordered around other
loads and stores.  A correct queue therefore requires tricky use of barrier 
and fence instructions. 

Moreover, given the number of descriptor queue implementations in a
typical OS, it is surprising and disappointing that generic
functionality cannot be shared among implementations, nor can
implementations \emph{compose} efficiently into a pipeline, in the
manner of more heavyweight data transfer frameworks like Streams in
AT\&T~\cite{Ritchie1984} and Plan 9~\cite{Presotto1993}, or the
protocol modules in the
x-Kernel~\cite{Hutchinson:1991:XAI:110809.110815}.

\subsection{CleanQ}

The duplication of code and continuing stream of new bugs in ad-hoc
descriptor queues led us to develop \System, a formally-specified data
transfer model for descriptor queues with an associated C-language
interface.

To the best of our knowledge, \System is the first practical,
formally-specified, general-purpose descriptor ring abstraction, and
we show that the generality and strict semantics of \System come with
negligible performance penalty compared with poorly-specified (but
well-implemented) subsystems in production use.

By decoupling rigorously defined data transfer semantics from
implementation, \System allows clients and implementations to be
developed and tested separately with much greater assurance of
correctness. \System is a specification and not an implementation,
thus leaving orthogonal issues like metadata, setup, and additional
semantics to the implementation while adhering to a single common
model for data transfer.

\System is highly general: it can express a variety of
hardware descriptor queues as well as communication channels between
processes in an OS.  
We demonstrate this functionality by later implementing, among other
things, a debugging module which is transparent to a \System queue but
applies rigorous online checking of its arguments.

\System is \emph{not} ``yet another queue implementation'', nor is it
a bug-finding technique for existing implementations.  Moreover, our
goal is not to build a formally verified system (such as seL4
\cite{Klein2009} or CertiKOS \cite{Gu2016}), but rather a sound basis
for reasoning about the system and its behavior and having clearly
defined semantics.  \new{\System is an example of how a useful subset
  of the benefits and guarantees of full-stack verification can be
  practically introduced into existing systems in a portable and
  incremental manner}

A concise, formally-sound model such as \System is
essential to the development and proof of formal systems software, but
it is just as important for non-verified systems such as Linux. As
long as the implementation of a \System module adheres to the
specification, we can guarantee properties proved on the formal model.

\section{Model}

\label{sec:design}

\begin{figure}
\includegraphics{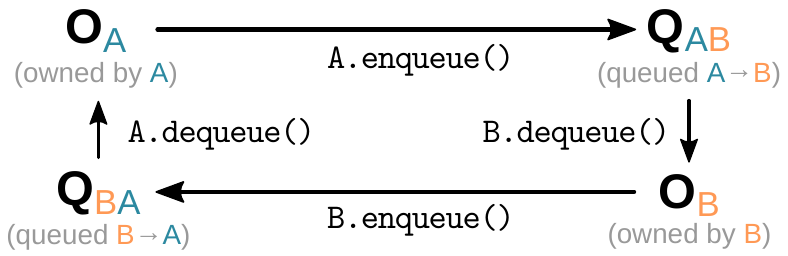}
\caption{\label{fig:ownership}Ownership transfer}
\end{figure}

The interfaces we consider (e.g.~\texttt{virtio}) all transfer data in
\emph{buffers} (packets, VM pages, disk blocks, etc.) between
\emph{processes} (including software processes, device drivers,
hardware devices, etc.).  The copy itself is simple (for a zero-copy
implementation, it is completely absent).  The principal difficulty is
the bookkeeping: When can a process safely read a buffer it has
received?  When must it stop writing before handing it off?  When,
exactly, \emph{is} the buffer handed off?

We therefore take the concept of \emph{ownership} as our primitive
abstraction, and base our formal invariants\footnote{Properties that
  must \emph{always} hold.} on the following four properties that must
hold if an entity can really be said to ``own'' a thing:

\begin{enumerate}
\item A thing has at most one owner.
\item If an entity owns a thing, it has exclusive use of it.
\item An entity knows whether it owns a thing or not.
\item Ownership can be transferred.
\end{enumerate}

From the first property we infer the fundamental invariant of the model:
\begin{displaymath}
O_A \cap O_B = \emptyset
\end{displaymath}
The set of things owned by $A$ ($O_A$) is disjoint from the set of things
owned by $B$ ($O_B$), for any processes $A$ and $B$.  Note that a process is
anything that might read or modify a buffer: a user-space process, a device
driver, a hardware component (e.g.~a network card).

The second property expresses the most important guarantees that the system must
provide to processes (and that processes must, in turn, respect):  First, if
$A$ owns a buffer, any changes to the buffer visible to $A$ must be due to
modifications $A$ itself made since gaining ownership;  Second, no other
process ($B$) may rely on the contents of the buffer until $A$ relinquishes
ownership and $B$ acquires it;  Third, all changes caused by $A$ (while it
owned the buffer) must be visible to $B$ \emph{immediately upon acquiring
ownership}.  This guarantees isolation among processes, and provides clear
requirements for any code needing to manage a weak or non-coherent memory
system (by dictating barriers, flushes, etc. see Section \ref{sec:fences}).

The third and fourth properties force us to elaborate the formal
model: If atomic transfer were possible, we could stick with just the
sets by $A$ and $B$ and the bookkeeping problem would be
straightforward.  Generally however, no atomic transfer of ownership
is possible: most implementations (especially hardware) transfer
buffers by means of a descriptor ring or similar.  The relinquishing
process enqueues a descriptor referring to the buffer to be
transferred, which the acquiring process (eventually)
dequeues.

\new{Note that the transfer sets ($Q_{AB}$ and $Q_{BA}$) at this
point do not preserve ordering. We add the FIFO property by refining
the model in section \ref{Refinement}.}

While a buffer is in the queue (descriptor ring),
it \emph{cannot} be said to belong to either $A$ or $B$ in a way compatible
with our 4 properties.  If the buffers queued from $A$ to $B$ ($Q_{AB}$) belong to
$A$, $A$ is free to modify them as it likes (property 2).  But as soon as the
descriptor is dequeued, $B$ will assume it owns it (e.g.~the NIC will start
writing).  As enqueue and dequeue are asynchronous, $A$ has no way of knowing
when to stop writing!  Likewise, assigning ownership of $Q_{AB}$ to $B$
violates property 3: $B$ gains ownership (and thus responsibility) without being
informed (when $A$ enqueues).  The queues are therefore distinct from the
ownership sets (and from each other):
\begin{align*}
O_A \cap Q_{AB} = O_B \cap Q_{AB} = \emptyset \\
O_A \cap Q_{BA} = O_B \cap Q_{BA} = \emptyset \\
Q_{AB} \cap Q_{BA} = \emptyset
\end{align*}

This is the complete model, illustrated by Figure \ref{fig:ownership}.  Here
we see the four sets describing the transfer of ownership between processes
$A$ and $B$, and the allowable transitions. Barring \texttt{A.register()} and
\texttt{B.register()} (which add buffers to, and remove them from
bookkeeping), the four operations (\texttt{A}|\texttt{B.enqueue()},
\texttt{A}|\texttt{B.dequeue()}) transfer ownership of buffers clockwise: $O_A
\rightarrow Q_{AB} \rightarrow O_B \rightarrow Q_{BA} \rightarrow O_A$.

One final invariant completes the model, and expresses that buffers are never
lost, or invented out of thin air:
\begin{displaymath}
O_A \cup Q_{AB} \cup Q_B \cup Q_{BA} = \textsc{const}
\end{displaymath}

\subsection{Modelling the Intel i82599 Descriptor Ring}

\begin{figure}
\includegraphics[width=\columnwidth]{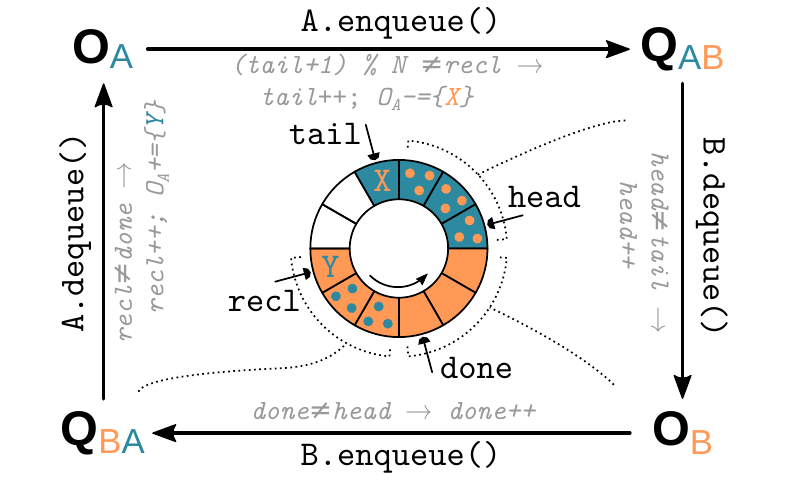}
\caption{\label{fig:ownership_rb}Ownership in the i82599 ring buffer}
\end{figure}

Figure \ref{fig:ownership_rb} illustrates the descriptor ring buffer of the
Intel i82599 10GbE network controller
\cite{IntelX520, intel82599_datasheet} used in Intel's popular X520
server network cards, and how it is
interpreted in the \System model.  The ring itself (figure center) is a
circular buffer with two pointers: \texttt{head} and \texttt{tail}.  The
descriptors from \texttt{head} up to (but not including) \texttt{tail} are
those enqueued but not yet taken by the device i.e.~the set $Q_{AB}$ (here $A$
is the driver and $B$ the NIC).

The driver enqueues $\texttt{X}$ by writing at \texttt{tail}, then
incrementing the pointer, atomically transferring $\texttt{X}$ from $O_A$ to
$Q_{AB}$.  The NIC dequeues by incrementing \texttt{head}, atomically moving a
buffer from $Q_{AB}$ to $O_B$, (\texttt{done} up to \texttt{head}).  The
\texttt{done}\footnote{There are two hardware modes: An \emph{explicit}
\texttt{done} pointer, or a \texttt{done} bit in the descriptor which defines
an \emph{implicit} done pointer} pointer is only modified by 
hardware and points to the last (oldest) buffer that the NIC has dequeued
but not yet processed.

Only \texttt{tail}, \texttt{head} and \texttt{done} have hardware-dictated
meaning --- The NIC doesn't distinguish (and doesn't need to) between buffers
that are enqueued back to the driver ($Q_{BA}$) and already dequeued in software
and ready for reuse (unshaded descriptors in~\autoref{fig:ownership_rb}).
The driver keeps track of which descriptors it has dequeued (and are safe for
reuse), with the \texttt{recl} pointer.  This points to the oldest descriptor
in $Q_{BA}$ (the last shaded). The i82599 processes buffers in order.
\texttt{recl} divides the region between \texttt{tail} and \texttt{done} into
the returned descriptor queue ($Q_{BA}$) (\texttt{recl} to \texttt{done}) and
`free' descriptors (\texttt{tail} to \texttt{recl}).

All four of the queue operations consist of atomically incrementing a pointer
(as indicated in gray in~\autoref{fig:ownership_rb}, together with the guards
($\rightarrow$) against letting the head of a queue overtake its
tail).

This shows that the \System specification and its notion of
ownership do, in fact, model the i82599 hardware queues -- \System
closely corresponds to the design of real, high-performance
hardware.  The extremely simple implementation possible in this case
also demonstrates that there is no inherent overhead to a
well-specified formal interface, such as \System.

$O_A$ (the buffers owned by $A$) cannot be defined by the content of the
descriptor ring.  $A$ might have e.g.~\texttt{register}-ed a pool of buffers,
shared between multiple queues.  Operations on $O_A$ are thus defined
abstractly:
\begin{align*}
\texttt{A.enqueue(X)}:\quad &O_A\mathop{:=} O_A-\{X\}\\
\texttt{A.dequeue(Y)}:\quad &O_A\mathop{:=} O_A\cup\{Y\}
\end{align*}
Any \emph{implementation} of \texttt{A.enqueue()} must cause $A$ to relinquish
ownership of $\texttt{X}$ and that of \texttt{A.dequeue()} cause $A$ to take
ownership of $Y$.  This is not a property on the ring buffer, but rather a
correctness requirement for software that \emph{uses} the ring buffer: It
tells the programmer exactly \emph{when} they must relinquish ownership, and
exactly when they may assume they have re-acquired it.

\subsection{Refinement} \label{Refinement}

\begin{figure}
\includegraphics{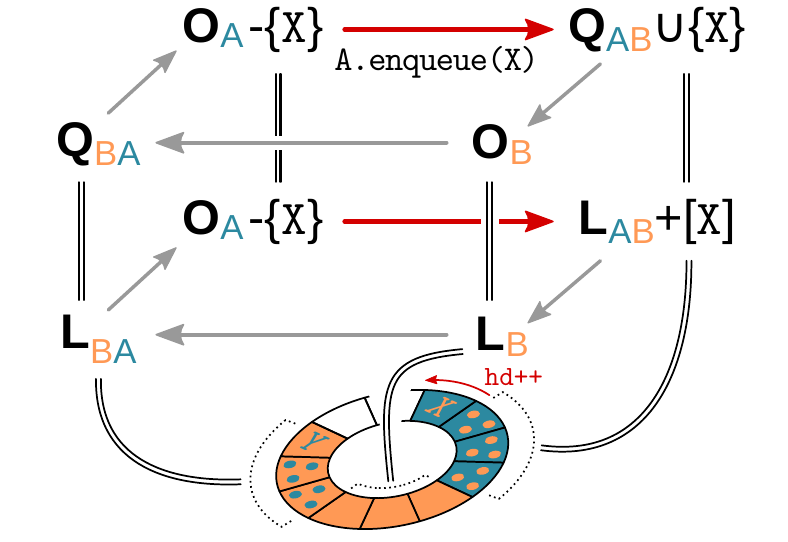}
\caption{\label{fig:refinement}Refinement steps}
\end{figure}

This notion of a `specification to be implemented' is a \emph{data refinement}
(as used e.g.~in the seL4 proof~\cite{Cock2008, Winwood_Mind_The_Gap}), and is
also how our i82599 interpretation is 
formally specified. \autoref{fig:refinement} depicts a stepwise refinement
of \texttt{A.enqueue(X)} from the abstract set-based model described so far,
via an intermediate model where queues become \emph{lists} (establishing FIFO
order), to the ring buffer model just described.

Each layer is the ownership transfer ring (c.f.~\autoref{fig:ownership}) at a
given refinement level.  Double lines indicate elements linked by the
\emph{state relation} e.g.~the set $Q_{AB}$ contains exactly the elements of
the \emph{list} $L_{AB}$ which is in turn the descriptors from \texttt{done}
up to \texttt{head} in the ring.  Red arrows highlight the refinement of
\texttt{A.enqueue(X)}, from set insertion ($Q_{AB} \cup \{\texttt{X}\}$) to
list append ($L_{AB} + [\texttt{X}]$) and finally pointer increment
(\texttt{head++}).

The state relations, and refined datatypes and operations are all formalised
in Isabelle/HOL.  Following the convention used in Formal Methods
conferences, we do not include them here for space reasons, but all 
Theory sources will be published and are available on request.

\subsection{Concurrency}
\label{sec:concurrency}

\System is fully concurrent, and mandates no locks.  $A$ and $B$ may
simultaneously enqueue and dequeue to their shared queues, as long as the
invariants are preserved.  For the i82599 this reflects that, for example,
\texttt{head} is updated by the NIC obliviously to everything except that it
does not overtake \texttt{tail}.  The driver is free to enqueue at
\texttt{tail} at precisely the moment that the NIC dequeues at \texttt{head}.

The \emph{strict postconditions} of~\autoref{fig:refinement} are not
preserved by the actions of a concurrent process.  For example, the strict
postcondition $Q_{AB}^\text{new} = Q_{AB}^\text{old} \cup \{\texttt{X}\}$ for
\texttt{A.enqueue(X)} is invalidated if $B$ dequeues $\texttt{X}$.

\begin{figure}
\includegraphics{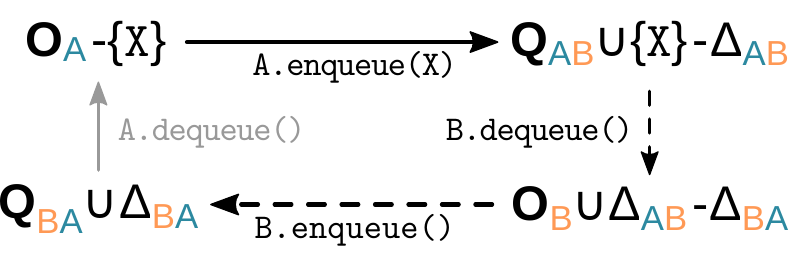}
\caption{\label{fig:frame}Weak postconditions for concurrency}
\end{figure}

In reasoning about $A$, we cannot rely on $X$ being in $Q_{AB}$, just because
$A$ has executed \texttt{enqueue(X)}.  We \emph{can}, however, infer that $X$
is in one of $Q_{AB}$, $O_B$ or $Q_{BA}$ --- everywhere $B$ might have put it,
without $A$ doing anything (i.e.~calling \texttt{dequeue}).
\autoref{fig:frame} summarizes the \emph{weakened} postcondition for
\texttt{enqueue(X)} that is preserved under \emph{interference} by $B$.

These weakened postconditions are a prerequisite for verifying the correctness
of a particular implementation under full concurrency, using 
Owicki-Gries~\cite{OG} logic as, for example, in the verification of the 
eChronos real-time operating
system~\cite{Andronick_LMMR_16}.  We have formalized these, also in
Isabelle/HOL, including noninterference and refinement proofs for all abstract
levels. 
\RH{Can we really call this a framework? Seems like reviewers confused this}

\subsection{Caches and Memory Fences}
\label{sec:fences}

Knowing exactly when ownership is gained and lost is essential to knowing
exactly which \emph{cache management operations} and
\emph{fences}/\emph{barriers} are needed, and when, in order to correctly
provide the guarantees implied by our `four properties of ownership'.  In
particular, weak-memory-model architectures (such as ARM and
Power~\cite{ArmMM}) and partially- or non-coherent systems (e.g. accelerators)
may violate the exclusivity guarantees by reordering memory operations past
the ownership transfer (by reordering or speculatively
executing instructions, or by serving stale values from non-coherent caches).

\begin{figure}
\includegraphics[width=\columnwidth]{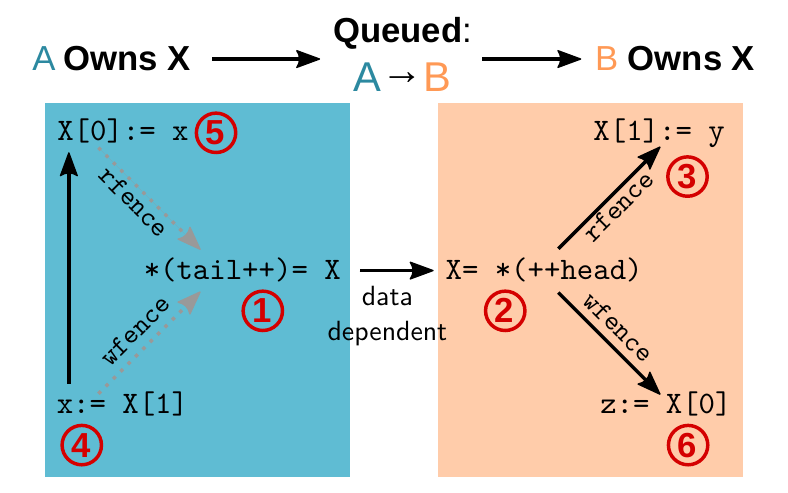}
\caption{\label{fig:weak_mem}Ownership implies fences}
\end{figure}

Consider~\autoref{fig:weak_mem}, depicting the transfer of the buffer
\texttt{X} between sender $A$ and receiver $B$, on a hypothetical
very-weak-memory architecture (similar situations are or were observable on
IBM Power and DEC Alpha systems).  In the absence of fences (barriers), the
only orderings guaranteed are those with a data dependency, marked with a
solid arrow.  The two dotted arrows between $A$'s modifications to the buffer
and its relinquishing ownership (incrementing \texttt{tail}) only indicate the
\emph{intended} ordering; These operations may occur in any order.  In
particular, the execution order indicated by the circled red numbers is
consistent with the constraints (and can actually be observed).

Here we see $A$ relinquish ownership (1), then $B$ acquire it (2) and
immediately write value \texttt{y} to \texttt{X[1]} (3).  Only \emph{then} are
$A$'s modifications scheduled: It reads the value in \texttt{X[1]}
(\texttt{y}, at 4), and writes it into \texttt{X[0]} (5).  Finally, $B$ reads
the value in \texttt{X[0]}, and sees the value \texttt{y} that it itself
wrote.  $B$ has communicated to itself by traveling into the `past'!

A sufficient fix in this case is to add the read and write fences as
indicated. $A$'s read from \texttt{X[0]} is forced to commit before the
increment to \texttt{tail} (as it also involves a read), and likewise the
write to \texttt{X[1]}.  The read and write fences in $B$ are redundant in
this example (as in fact is $A$'s write fence due to the data dependency).  If
not all instructions in $A$ and $B$ are known however, all four barriers are
necessary.

Where the ownership model helps here is that all four barriers can be inferred
from the guarantees and responsibilities of ownership: $A$ must ensure that
any writes to \texttt{X} become visible to $B$ \emph{before} $B$ learns that
it owns \texttt{X} (i.e. $A$'s write to \texttt{tail} becomes visible to $B$).
The \emph{last} point at which $A$ can ensure this (as $B$'s dequeue is
asynchronous) is when it enqueues \texttt{X} -- Hence the write fence before
updating \texttt{tail}.  Likewise $A$ must not rely on \texttt{X} after
relinquishing ownership -- Hence the read fence before the enqueue.  An
equivalent argument implies the necessity of the fences on $B$'s side (e.g.~in
the absence of a data dependency).

Furthermore, it should be possible to \emph{automatically} place the required
fences, for some combination of a (possibly-stronger) memory model (e.g.~ARM
or TSO) and known code in $A$ and $B$.  Such automatic inference in code of
similar complexity was demonstrated by Liu et.~al.~\cite{Liu:2012:Fences}

\section{Interface and Implementation}
\label{sec:implementation}
\label{sec:impl:cinterface}
\label{ssec:semantics}

In this section we describe the C interface we derive from the formal
model in the previous section, together with a set of implementations
(termed ``modules'', following Unix Streams~\cite{Ritchie1984}) we
have built and evaluated for inter-process communication and device
drivers.

\begin{figure}[!h]
	{\footnotesize 
		\begin{tikzpicture}[node distance=0.5em, auto]
		\node[draw,minimum height=2em,minimum width=27.5em] (generic)  {\System 
			Generic 
			Interface};
		\node[draw,below=0.1em of generic.south, minimum height=2em,minimum 
		width=27.5em] (common)  {Common Code};
		\node[draw,below=1.1em of common.south west, minimum height=2em,minimum 
		width=4em,anchor=west] 
		(module1) {NetworkQ};
		\node[draw,right=0.5em of module1,minimum height=2em,minimum 
		width=4em,anchor=west] 
		(module2) {AHCIQ};
		\node[draw,right=0.5em of module2,minimum height=2em,minimum width=4em] 
		(module3) {LoopbackQ};
		\node[draw,right=0.5em of module3,minimum height=2em,minimum width=4em] 
		(module4) {DebugQ};
		\node[draw,right=0.5em of module4,minimum height=2em,minimum width=4em] 
		(module5) {EthernetQ};
    \node[draw,below=0.1em of module3,minimum height=2em,minimum width=4em] 
    (module7) {IPQ};   
    \node[draw,left=0.1em of module7,minimum height=2em,minimum width=4em] 
    (module6) {UDPQ};
    \node[draw,right=0.1em of module7,minimum height=2em,minimum width=4em] 
    (module8) {DPDKQ};
		\end{tikzpicture}
	}
	\caption{Implemented modules of \System Library}
	\label{fig:overview}
\end{figure}
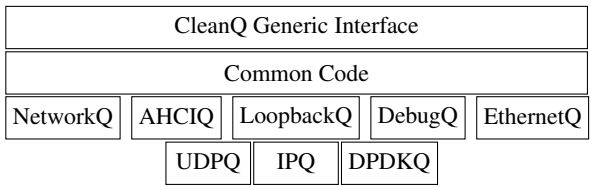

Figure~\ref{fig:overview} shows the software architecture.  
To show the applicability of \System 
to different devices and other use-cases, we have implemented modules
for an AHCI~\cite{AHCI} storage host adapter, an Intel e1000
NIC~\cite{intel631_manual}, an Intel i82599 10Gb
NIC~\cite{intel82599_datasheet}, a Solarflare SFN5122F low-latency
NIC~\cite{solarflare_sfn5122f}, a network protocol stack implementing
UDP/IP, a shared-memory inter-process queue, and a DPDK~\cite{DPDK} module for the Intel 
i82599 10Gb NIC. We also implemented a debug module, which checks the interface 
contract at runtime. We describe the detail of the network stack and debug 
modules in~\autoref{sec:impl:modules}. 

We have applied \System in Linux, DPDK, and a microkernel-based
research OS, showing that it is deployable across multiple, complete
existing systems.  An additional, Rust-based implementation which
exploit's Rust's ownership-based type system is beyond the scope of
this paper. 

\subsection{Definition}

\autoref{fig:cleanqinterfaceX} shows the C
declarations for \system.  This interface is implemented by generic
code which performs various integrity checks (such as region bounds
for buffers) before calling corresponding module-specific methods in a
vtable associated with the \texttt{struct cleanq} argument. 

The \texttt{enqueue} and \texttt{dequeue} methods must adhere to their
specification introduced in~\autoref{sec:design}.   The additional 
\texttt{notify}, \texttt{register}, and \texttt{deregister} calls are
described below.

\begin{figure*}[!h]
    \begin{footnotesize}
        \begin{lstlisting}[style=code]
err_t cleanq_register(struct cleanq *q, void *mem, regionid_t* rid);
err_t cleanq_deregister(struct cleanq *q, void *mem, regionid_t rid);
err_t cleanq_enqueue(struct cleanq *q, regionid_t rid, size_t offset, size_t length, 
	             size_t valid_data, size_t valid_length,uint64_t flags);
err_t cleanq_dequeue(struct cleanq *q, regionid_t* rid, size_t* offset, size_t* length, 
	             size_t* valid_data, size_t* valid_length, uint64_t* flags);
err_t cleanq_notify(struct cleanq *q);
    \end{lstlisting}
    \vspace{-3mm}
    \caption{The \system library interface}
    \label{fig:cleanqinterfaceX} 
    \end{footnotesize}
\end{figure*}



We now describe the semantics of the \System interface functions in detail.
Interface calls that do not satisfy the required preconditions are
bugs on the caller side and cause undefined behavior (though many are
caught by the generic checking code).  As in~\autoref{sec:design},
``process'' denotes anything that changes buffers, for instance a
software driver or the hardware of a network interface card. 

\textbf{Creation and destruction} of queues is not part of the interface,
since these processes are highly implementation specific and typically
need module specific parameters, such as device registers of a network
card, a shared memory buffer for the loopback/IPC queue, or another
queue in case of the debug queue.  Creating a queue must include initializing
the \texttt{cleanq} data structure including the generic state and
vtable. 

\textbf{Register} takes a contiguous region of memory, previously
owned by neither side of the queue, inserts it into the set of owned
buffers, and returns an identifier for it to be used in subsequent
enqueue and dequeue operations.  Register is typically used in
conjunction with buffer pools or slab allocators that allocate a large
chunk of memory at once.

Register is of practical importance in cases where address-related
state must be set up in advance, for example regions for RDMA-based
transfers, or programming an IOMMU to make a region of memory
accessible to both sides of the queue.  In simple shared-memory cases
it can be implemented as a null operation which returns the pointer
address as the handle.

The \texttt{mem} argument is an OS-specific handle to a memory
resource, e.g. a pointer to anonymous memory, a file handle to a
mapped segment, or a capability to physical memory. 
The memory region has to be at least read-accessible from the calling
process.  At no time may the set of registered regions overlap
with each other.

\textbf{Deregister} removes a previously-registered region with
the supplied id from the queue.  Deregister can only succeed if
the region has not already been deregistered, \emph{and} all memory in
the region is currently owned by the calling process. 

\textbf{Enqueue} enqueues a buffer of a previously registered region for ownership transfer. 
Buffers are identified by a region id, an offset into this region and a 
length. The buffer described by offset and length, must lie within the
registered region, and must be owned by the process (i.e. a buffer 
cannot be enqueued twice without dequeuing it beforehand).
The operation can fail if the underlying queue has run out of space.

The valid payload is specified by a further offset and
length within the buffer, allowing clients to leave space for headers
and footers (meta data) added later. 

As specified, a successful enqueue
relinquishes ownership of the buffer and inserts it into the transfer
set. Eventually the ownership of the buffer will be obtained by the
peer process, but there is no guarantee when this happens.

A client must not alter a buffer once it has given up ownership, and
doing so will result in undefined behavior.  Since we know precisely
when we yield the ownership and which memory region is described by
the buffer, the implementation can and must guarantee that all changes to the 
buffer are observable (using memory fences) before the ownership is transferred.

\RH{This seems to confuse more than help: Post-modification of the 
	buffer by the client might even be detectable by static analysis
	 (though we do not do this)}

The \texttt{flags} field allows additional metadata to be passed along
orthogonally with the buffer, with the proviso that the formal
semantics from~\autoref{sec:design} are not altered in any way.  For
example, a DMA copy engine might require the client to distinguish
source and destination buffers for a copy.  

\textbf{Dequeue} removes a previously enqueued buffer from the
queue and transfers ownership of the buffer to the calling process. As
long as the process owns a buffer, the process can alter the contents
of this buffer. Dequeue can be called any time but returns an error
if there is nothing to dequeue. 

Absent an error, a correct implementation must return a valid buffer, i.e. one that
is within a previously registered region and is 
not yet owned by the calling process.  As with enqueue, a subset of
the buffer can be declared valid using \texttt{valid\_data} and
\texttt{valid\_length}, to allow stripping of headers (for example, in
the UDP queue example we show in ~\autoref{sec:impl:modules}). 

Metadata about the transfer can, as with enqueue, be returned in
\texttt{flags}.  Again, this is implementation defined but must be
orthogonal to the memory ownership semantics.  It can be used to
signal corrupt packets from a network adapter, for example, or as part
of a chaining protocol (\autoref{subsec:discussion})

\textbf{Notify} is an optional performance optimization
mechanism: for example, a doorbell informing the process on the
other side of the queue that there \emph{might} (i.e. no guarantee) be buffers in the queue that
are ready for processing.  It has no formal semantics at all, and its
use (or omission) must not affect the correctness of any
implementation relative to the specification. 

\subsection{Module Composition and Debugging}
\label{sec:impl:modules}

We have implemented \System modules which communicate with a variety
of hardware devices using their native descriptor format and protocol,
along with inter-process communication channels which pass descriptors
in shared memory using a variant of
FastForward~\cite{Giacomoni2008}. Despite incorporating basic runtime
checks, we show in Section~\ref{sec:evaluation} that these modules are
comparable in performance with the ``native'' implementations they
replace. However, \System's formally specified interface has a further
advantage: in contrast to ad-hoc, C-specified queues, \System modules
can compose in a pipeline or stack, analogous to System V streams.
The ``null'' implementation (a module which sits in front of another
\System module but simply passes data through) imposes negligible
overhead, and we have implemented a \emph{debug module} which augments
\System's default bounds checks with more extensive bookkeeping to
detect violations of the queue's contract by either client or
downstream module.  For example, it maintains an operation log for
debugging purposes, and detects overlapping or duplicate enqueues,
which prevent ``double fetch'' race condition
vulnerabilities~\cite{Wang2017}.

\subsection{Networking}

We have built a full-duplex UDP protocol stack which sits atop
a \System module implementing a NIC's hardware queues and which itself
consists of two \System modules: one for the UDP headers and one for
IP and Ethernet headers.  The result is a dataplane
implementation similar to Arrakis~\cite{Peter2014}.
The structure of the enqueue call is shown in figure~\ref{fig:udp_stack}.
\begin{figure}[h]
\begin{footnotesize}
\begin{lstlisting}[style=code]
struct udp_q* que = ... 
struct cleanq* q = (struct cleanq*) que;
cleanq_enqueue(q, ...){
  // Some interface checks	
  q.enq(q, ...){
    // Build UDP header
    q->ip_q.enq(q->ip_q, ...){
      // Build IP + Ethernet header
      ip_q->nic_q.enq(ip_q->nic_q, ...){
        // Build descriptor and inform Hardware
} } } }
\end{lstlisting}
\end{footnotesize}
\vspace{-4mm}
\caption{Stacking queues to implement UDP}
\label{fig:udp_stack}
\end{figure}  
In order to implement different layers of the stack, we use
\texttt{valid\_data} and \texttt{valid\_length}.  When receiving a packet,
each layer reads and interprets the header found at offset
\texttt{valid\_data}.  To pass a packet up to the next higher layer
\texttt{valid\_data} is incremented by the header size, such that the next
higher layer will ignore the current layer's header.

\RH{Might talk about sk\_buffs and other mechanisms here? We mention having some
	space before the buffer etc.}

\subsection{Discussion}
\label{subsec:discussion}
\RH{Move this to evaluation (Suggested by Reviewer)?}
Our experience building a number of \System modules, and composing
them, has so far been very positive.  Implementation is
generally straightforward, similar to a Virtio queue or an ad-hoc
implementation, and establishes that the model is sufficiently general
to cover all the use-cases we have encountered so far.

\System's formal semantics make it very clear what obligations
exist for a module programmer at every point in the code, and remove
most of the uncertainty about what the code needs to guarantee and
when.  The use of stackable modules provides the expected benefits in
composability, and we have made extensive use of the debug module for
checking.

Compared with other queue implementations in systems like Linux,
however, \System is something of a radical simplification, and this
might raise several concerns.

Firstly, we are paying the price of abstraction: the clearer interface
requires indirect method calls for each module.   Historically these
have been viewed as expensive, but as we show in
Section~\ref{sec:evaluation} modern processors have reduced this
overhead to considerably less than the cost of, e.g., formatting hardware
descriptors, and so this appears not to be an issue.

Secondly, each enqueue or dequeue operation acts on a single buffer:
there is no batching.  In practice, the cost of multiple
enqueue/dequeue operations is sufficiently small in our
implementations that this does not degrade performance significantly.

Finally, we do not directly support chaining of multiple,
discontiguous buffers as with BSD \texttt{mbuf}s or Linux
\texttt{sk\_buf}s.  Instead, we chain buffers using a simple protocol
\emph{above} \System's single-buffer enqueue/dequeue operations.  As
with batching, the additional overhead is small for our usecases. 

Our argument, backed up by performance measurements, is that
the simplicity and rigorous semantics of \System outweigh the small
overhead the design might incur.

\newcommand{\reps}{100,000 }
\section{Evaluation}
\label{sec:evaluation}

To evaluate the performance of \System we first benchmark the 
overhead of our implementation of the interface, then compare the equivalent operations 
of Virtio to our queues. Following this, we set the overhead into perspective 
of a real application. We then further evaluate the different mechanisms of 
stacking, the debug queue and finish the performance benchmarks with a more 
complex example of an implementation of a UDP stack based on our queues as 
well as on DPDK. Finally we discuss the performance of \System based on the 
previously presented benchmarks.

With these benchmarks we show that there is no significant performance loss
when changing from existing systems to \System while we gain
a clean, easier to use, well-defined interface with the ability to stack 
queues. Furthermore, in our implementation of the interface we added
sanity checks on the buffers through the thin library layer that are 
not included in most systems.

All experiments were conducted on a two-socket Intel Xeon E5-2670 v2
(Ivy-Bridge, 2.5 GHz) system with hyper threading disabled.  We used
an i82559-based Intel X520 dual-port 10GbE card to evaluate the
performance of the UDP queue. Unless indicated otherwise, all
measurements are taken using the timestamp counter of the
processor. We evaluated \System on Linux (Ubuntu 18.04 LTS) and a
microkernel-based research OS.

\subsection{The Overhead of Common Code}
\label{ssec:overhead_library}

This benchmark shows that the performance overhead of our C implementation,
which provides some sanity checks as common code, is 
small in absolute terms for all four operations \emph{enqueue, dequeue, register} 
and \emph{deregister}.

The benchmark setup is as follows: We configured \System to use the 
\emph{loopback}-module which resembles an in-memory ring buffer where enqueue 
writes the descriptor into memory and dequeue reads the descriptor contents 
from memory and the corresponding pointers are updated accordingly. We measure 
at two points: at the calls to the interface and the calls to the 
module (before the vtable invocation). We run the benchmark of \reps 
repetitions and account for our measuring instrumentation.

\begin{figure}
	\center
	\includegraphics[width=\linewidth]{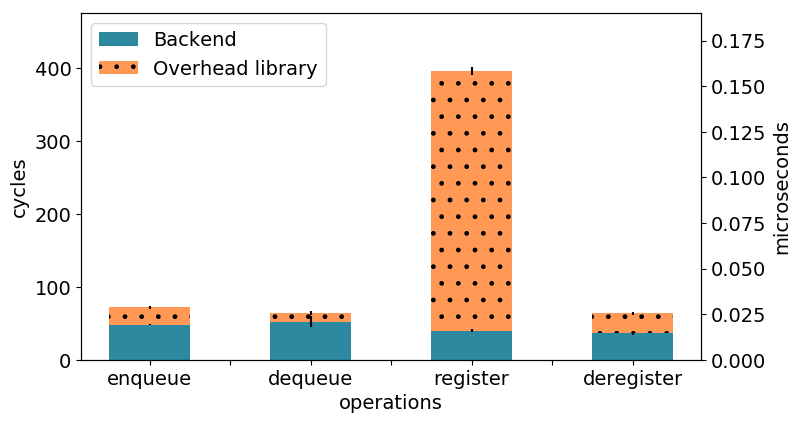}
	\caption{Overhead of C interface implementation}
	\label{fig:overhead_library}
\end{figure}
\autoref{fig:overhead_library} shows the median (and standard deviation) of 
each of the four operations. We observe that the cost of the thin library layer
of our \System implementation is on the order of tens of cycles 
for the enqueue, dequeue and deregister operation whereas register requires 
an additional check that the memory region is actually owned by the 
caller resulting in about 400 cycles overhead for a system call and 
the required bookkeeping.

The results show that our implementation adds little overhead 
in exchange for a well-defined and clean interface based on a formal model. The overheads 
of the fast-path operations enqueue/dequeue are less than 30 cycles and require
fewer cycles than the simple loopback module. We expect the register/deregister 
operations to be on the slow-path but nevertheless they 
only add a few hundred cycles at most for the bookkeeping operations. 
\\
\subsection{Comparison with Virtio}
In this benchmark we compare the operations of \System (with our loopback 
module) that have an equivalent 
in Virtio (add/get vs. enqueue/dequeue) to show that the performance is 
comparable. 

\begin{figure}

	\center
	\includegraphics[width=\linewidth]{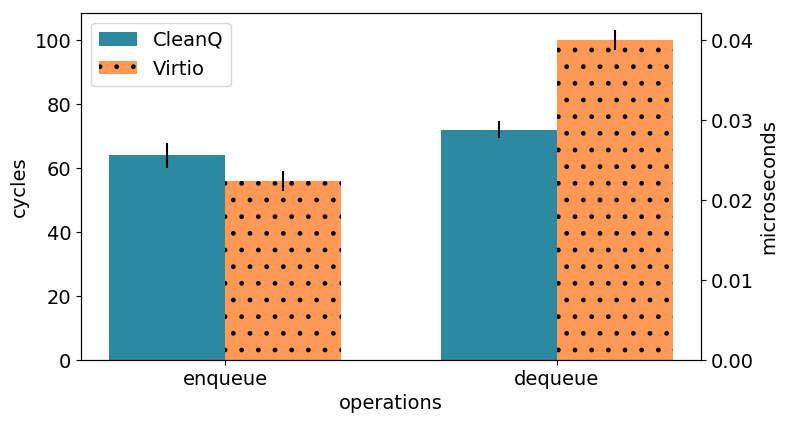}
	\caption{Performance of Virtio ring buffer (add/get) compared to \System 
		(enqueue/dequeue) both running on Linux}
	\label{fig:virtio_queue_stacked}
\end{figure}

\label{ssec:comparison_virtio}
We compare \System with Virtio (both on Linux) by measuring the calls to the application 
interface of \System and Virtio's virtqueue implementation. To measure the 
performance of Virtio we adapted one 
of the Linux Virtio tests, adding measurement code and increasing the number of 
repetitions. The Virtio test uses \texttt{virtqueue\_add\_inbuf()} to add 
buffers to the queue and after the host side has removed them, the buffers are 
reclaimed from the guest side by calling \texttt{virtqueue\_get\_buf()}.
Note, the host side of the virtqueue is accessible through a \emph{different} 
interface that requires a \texttt{memcpy} for adding data to the queue. 
For fairness we did not include the host side interface operations 
in this benchmark. The result of the benchmark is shown 
in~\autoref{fig:virtio_queue_stacked}. 
Enqueueing a descriptor to the Virtio virtqueue costs 56 cycles 
while enqueuing a buffer through our interface and then processing it in the 
module costs 72 cycles. Getting a descriptor 
from the virtqueue is more expensive at 100 cycles while dequeueing a buffer 
from \System only costs 64 cycles. 

Overall the performance is similar to Virtio's guest side while \System
provides additional checks on the buffers, a cleaner and simpler interface 
that allows for more complex constructs by stacking queues on top of each other.

\subsection{Placing library overhead in context}
To put our C implementation overhead of \System into perspective, we measure the total 
processing time of Memcached (v1.5.10)~\cite{Fitzpatrick2004} including network 
stack and hashtable lookup for set and get requests. This is a simple 
application context in which \system interface can be used.


We send small get/set requests (key + value < 16 bytes) over the network to
our Memcached server. We profile incoming requests by measuring the network 
stack processing time and the duration of Memcached handling the get/set 
request. Note, the resolution of the software timestamps provided by the network 
stack is one microsecond (or 2500 cycles).

\begin{figure}
	\center
	\includegraphics[width=0.9\linewidth]{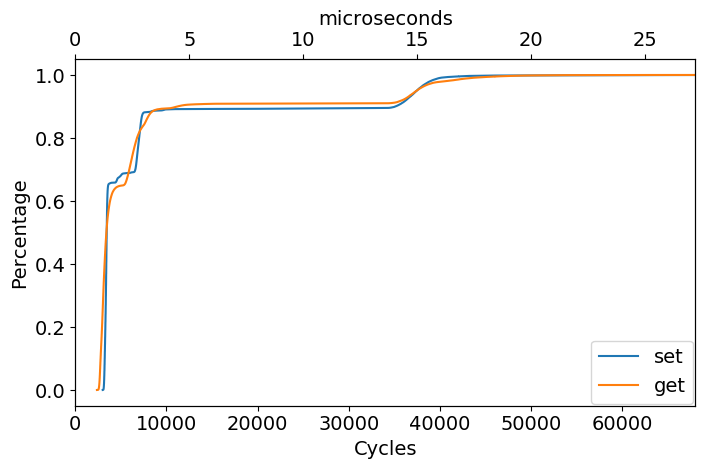}
	\caption{Memcached set/get processing time}
	\label{fig:memcached_process}
\end{figure}
\begin{figure}
	\center
	\includegraphics[width=0.9\linewidth]{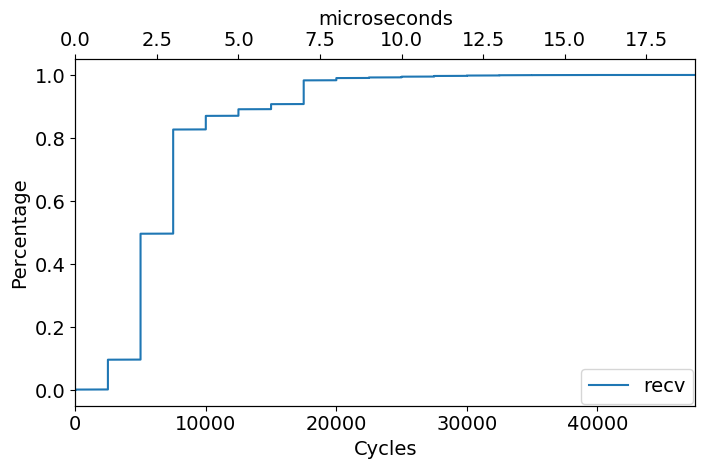}
	\caption{Linux: kernel to userspace network stack performance on receive 
		path}
	\label{fig:memcached_linux}
\end{figure}

The results, in the form of a CDF plot of \reps set/get 
operations, are shown in~\autoref{fig:memcached_process} for request handling in 
Memcached and~\autoref{fig:memcached_linux} for processing the packet in the 
network stack. The median time spent from the kernel to the userspace 
application on the receive path of a UDP packet is 3 microseconds or
around 7500 cycles. The median of both set/get is around 3450 cycles (1.3 
$\mu$s). Combining these two measurements results in 10950 cycles (4.38 $\mu$s) 
that are spent over the application's path on which the \system interface could 
realistically be used.

Comparing the time spent in the library on the fast path to the application's 
time spent in other code, leaves the overhead of the library at < 1\%. The 
small overhead of the library is dominated by the processing time of
other parts of the code. 

\subsection{Module Stacking Overhead}
\label{ssec:overhead_stacking}

In this experiment we measure the scalability of the implementations module
stacking.
We repeat the same experiment of~\autoref{ssec:overhead_library}, but we now 
stack ten \emph{null} modules on top of the loopback module. The \emph{null} 
module mimics a no-op: it only invokes the same operation on the next module in 
the stack and therefore all observed overhead originates from stacking itself. 
We measure at different levels of the stack the time it takes until the lower 
level completes the operation. Again, we conducted \reps runs for each 
operation. 

\begin{figure}[h]
  \center
  \includegraphics[width=\linewidth]{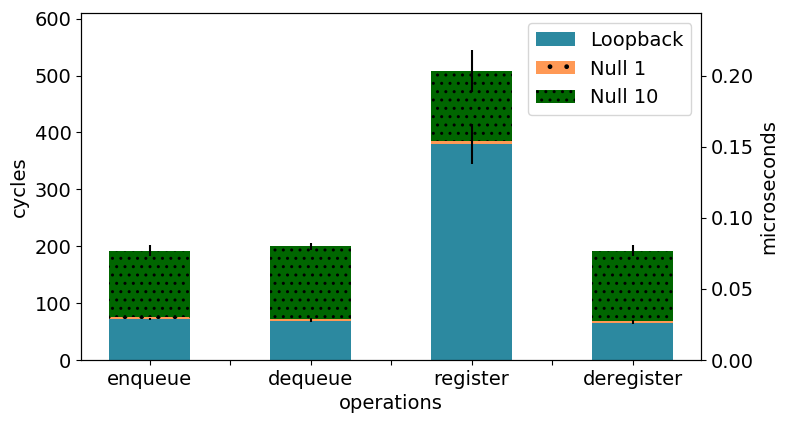}
  \caption{Overhead of stacking queues}
  \label{fig:overhead_stacking}
\end{figure}

\autoref{fig:overhead_stacking} shows the median execution time and standard 
deviation at three different points of measurement: \ei the base line 
(\emph{loopback}) represents the lowest level of the stack which includes only 
the loopback module (corresponding to \autoref{ssec:overhead_library})
\eii \emph{Null 1} represents the time taken when a single null module is 
stacked on top of the loopback module. We observe a negligible overhead for 
a single stack of less than 10 cycles for any of the operations. \eiii 
\emph{Null 10} measures the full stack of ten null modules stacked on top of 
the loopback module where the entire stack of ten modules results in about 100 
cycles overhead compared to the baseline.

Each additional module stacked on top corresponds to an additional indirect 
function call, which results in the overhead of less than 10 cycles for 
stacking a single module. Moreover, our results suggest that the overhead per 
stacked module stays constant when more modules are stacked. 
With this experiment we have shown that \system's stacking functionality is 
efficient.

\subsection{Debug Module Overhead}
\label{ssec:overhead_debug}

In this experiment we measure the overhead of the debug module that performs 
additional checks of buffer ownership on every queue operation. 

We repeat the experiment of~\autoref{ssec:overhead_library} with 
the only difference being that we stack the debug module on top of the loopback 
module. Again we perform \reps repetitions and measure the total execution time 
of each module in the stack.


\begin{figure}[h]

    \center
    \includegraphics[width=\linewidth]{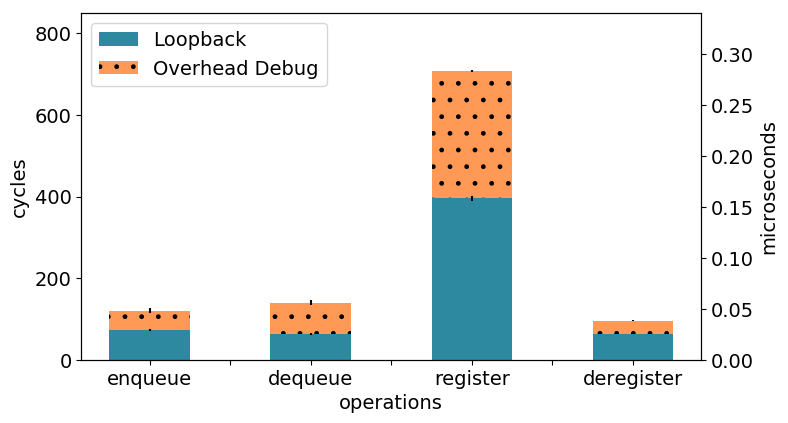}
    \caption{Overhead of debug queue}
    \label{fig:overhead_debug}
    \vspace{-2mm}
\end{figure}

\autoref{fig:overhead_debug} shows the median completion time for each 
operation including standard errors. We observe an overhead for the additional 
checks and stacking of the debug module of about 50-80 cycles for the enqueue 
and dequeue operations respectively and a total duration of 120-130 cycles. The 
deregister operation only adds about 20 cycles of overhead. Register is 
the most expensive operation adding 300 cycles.

The results show that even with tracking ownership, which requires lookup and 
updating internal data structures to reflect the change of ownership, we 
observe a total completion time for the two fast-path operations of less than 
130 cycles. Deregister simply checks whether the region has been registered 
before and then removes it if all buffers are owned by the caller, resulting in 
low overhead. The register operation requires verifying the size and access 
rights which requires a syscall in our implementation, resulting in a 2x 
increase which is, however, still less that $0.3\mu s$ 

Putting the overhead of the debug into perspective, despite an increase of up 
to 2x relative to the loopback module, the additional 50-80 cycles are dwarfed 
by the 3450 cycles processing time of our simple example application 
(Memcached) resulting in less than 2\% overhead.

\subsection{UDP Queue}
\label{ssec:udp_queue}

In this benchmark we show that we can implement a more complex construct 
based on our stacking mechanism to realize a high-performance, low 
overhead UDP network stack similar to that in the Arrakis
system~\cite{Peter2014}. This benchmark was implemented on the microkernel 
OS. 

This benchmark consists of a UDP/IP echo server using \System: a UDP module and an 
IP/Ethernet module both stacked on top of the e10k module which drives the Intel X520 dual-port 
10GbE card. The resulting queue is a stack of three modules. 
The network card has a distinct queue for transmit and receive and for each of 
the two hardware queues we initialize a \System stack. Note, the e10k module 
will need to convert from and to the descriptor format the network card 
understands. We generate 64-byte UDP packets and send them to the echo server 
with the \System stack. We measure the processing time for sending and 
receiving the packet on the echo server. 

\begin{figure}
  \center
  \includegraphics[width=\linewidth]{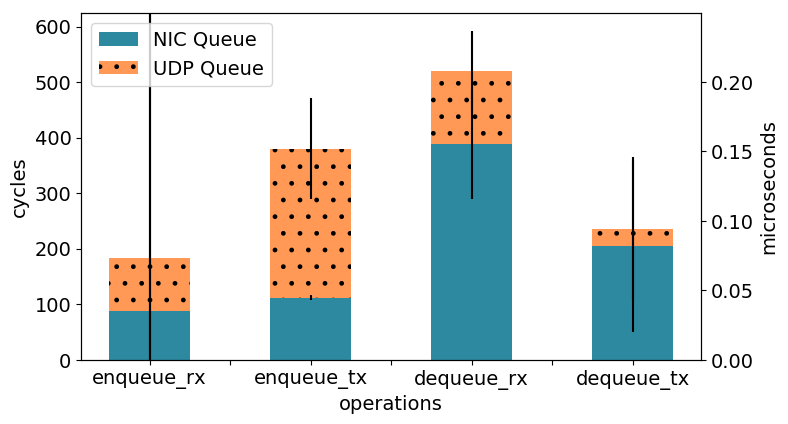}
  \caption{Performance of UDP queue}
  \label{fig:udp_queue}
  \vspace{-2mm}
\end{figure}

\autoref{fig:udp_queue} shows the median and standard deviation based on \reps 
measured packets for the e10k module and the rest of the UDP network stack 
(Ethernet/IP/UDP) combined. We observe a much higher standard deviation 
compared to previous experiments.

On investigation, this latency distributed is heavily bimodal: the
latency highly depends on whether the NIC hardware registers need to be 
updated. A write to the device register results in a 10x increase of the 
enqueue operation, but is only performed for a small fraction of enqueues.
How frequently this expensive register write occurs depends on 
load and batching heuristics, but is inherent in the hardware and not a feature 
of \System \emph{per se}. 

Enqueuing buffers into the transmit and receive queues of the NIC take about 90 cycles
whereas processing the UDP module takes about 100-250 cycles. When a buffer is 
written into the hardware descriptor queue, the descriptor needs to be 
formatted which takes most of the 90 cycles. The work done by the UDP module 
includes growing the valid pointers of the buffer to make space for the network 
headers as well as formatting the UDP, IP and Ethernet headers including 
the generation of checksums which account for the majority of the 250 cycles 
latency.

Dequeuing from the NIC queue is generally more expensive than enqueuing 
descriptors which result in 200-400 cycles latency. Dequeuing on the UDP module 
takes 20 cycles on the transmit queue and 120 cycles on the receive path.
Whenever the NIC completes a descriptor it updates the status bit of the 
descriptor which in turn ends up in main memory or the last level cache introducing
latency (90-150 cycles) when the descriptor is 
read by software. Moreover, the UDP module needs to verify the headers on the 
received descriptors.


To summarize, the performance characteristics of the hardware descriptor queues 
is generally dominated by the cost of formatting a descriptor and updating the 
register containing the receive and send pointers. Moreover, writing
hardware registers to inform the card about the software 
state can be expensive (> 3000 cycles) and batching the updates can amortize 
the cost which is the source of the large variance in this experiment.
We conducted similar experiments on a SolarFlare SNF5122F with comparable results.

Putting this into comparison of the library overhead, formatting a descriptor 
costs twice as much as our interface abstraction.

\subsection{DPDK}

In this benchmark we want to show that the \System interface can also be
integrated into existing systems without degrading performance. 

We implemented a module based on the DPDK Intel ixgbe driver for the
(i82599-based) Intel X520. We reuse the setup code and control plane of DPDK
\cite{DPDK} but reimplement the dataplane as a \System module. Additionally,
we stacked the UDP stack on top of the NIC \System module.

We compare the \System module (and the UDP stack) to the original DPDK driver, 
measuring the packets per second of a single NIC queue using one core running a UDP echo server
implemented using DPDK routines (send/recv burst). Furthermore,
we also measure the performance of the \System UDP stack directly using the 
\System interface. To generate load we implemented a benchmark using standard
sockets. On each core we run a thread which sends and receives the UDP packets
in a closed loop with a configurable amount of packets in flight. 
Table~\ref{tab:eval:dpdk} shows packets per second (median of 10 runs)
using minimum-sized packets.

\begin{table}
  \begin{center}
  	\begin{tabular}{ |c|c|c| } 
  		\hline
  		 & Pkt/s & Standard deviation \\
  		 \hline \hline
  		 DPDK & 705,000 & 3362 \\
  		 \System & 713,100 & 7305 \\
  		 \System UDP & 742,600 & 5066 \\
  		\hline
  	\end{tabular}
  \end{center}
\vspace{-3mm}
  \caption{DPDK vs. \System vs. \System UDP results}
  \label{tab:eval:dpdk}
  \vspace{-2mm}
\end{table}

\new{DPDK alone achieves a throughput of 705,000 pkts/s while
DPDK using \System achieved 713,100 pkts/s and the full UDP stack
using \System reached 742,600 pkts/s. Incorporating the \System 
interface into DPDK resulted in similar performance while
implementing a UDP stack using only \System increased performance by 5\%.}

This result shows that incorporating a \System into a high-performance networking
framework (DPDK) does not degrade throughput and can
even deliver better performance.

\subsection{Discussion}

In the evaluation we demonstrated that \System provides a clean and well defined 
queue abstraction with a strong notion of ownership transfer while still being 
lightweight \new{(compared to full stack verification)}
and able to deliver comparable performance to Virtio's virtqueue in 
a direct comparison. We have shown a C implementation of \System with low overhead 
in absolute numbers as well as when used in an application context the resulting overhead to be 
less than 1\% of the receive and processing time of Memchached get/set request. 
We further show the that the overhead of \System is not only dwarfed by 
application processing time but also by interfacing with the hardware itself 
such as by formatting descriptors and writing registers.

Furthermore, we have demonstrated \System's flexibility in building efficient 
protocols and adding strict bounds checks by stacking of modules -- a feature 
which is enabled by the well defined abstraction of \System. We have shown that 
stacking a module has a small overhead and is scalable to multiple modules. 
Finally, we have demonstrated this functionality by implementing a UDP network stack.


\section{Conclusion}
\label{sec:conclusion}

\System demonstrates that it is possible to unify many of the proliferation of
descriptor queues in common usage behind a single interface with strict formal
semantics, at no performance cost on the dataplane.

The benefits begin with the elimination of many subtle bugs which appear (and
reappear) whenever such an interface is defined informally.  These include
failing to catch all the usage cases, as well
as different interpretations of the interface between client and
implementation (which a strict formal specification prevents).  The
ownership-transfer model of \System is sound under lock-free concurrency, and
provides a framework for the verification of implementations.

The many \System modules already implemented show that such an interface is
widely applicable within an OS, and also permits the composition of modules to
provide reuse of functionality.

Moreover, this generality, composability, and soundness come at a low cost:
\System matches Virtio for operation latency and imposes less than 1\%
overhead on end-to-end application workloads such as Memcached.

The Isabelle/HOL formalisations will be published separately as an extended
technical report, and all \System modules and support code will be released
under an open source license.

%
%

\RA{THIS SHOULD BE AT 12 PAGES!!!}
\bibliographystyle{plain}
\bibliography{references}
\end{document}